\documentclass[12pt]{article}
\usepackage{amsfonts,bm}
\usepackage{graphicx}
\baselineskip
\offinterlineskip
\begin{document}

\begin{center}
{\bf Pauli Lubanski Vector Operator, Eigenvalue Problem 
and Entanglement}
\end{center}

\begin{center}
{\bf Garreth Kemp$^\dag$ and
Willi-Hans Steeb$^\dag$} \\[2ex]

$^\dag$
International School for Scientific Computing, \\
University of Johannesburg, Auckland Park 2006, South Africa, \\
e-mail: {\tt garrethkemp@gmail.com}\\
e-mail: {\tt steebwilli@gmail.com}\\
\end{center}

\strut\hfill

{\bf Abstract.} We study an eigenvalue problem for a
spin matrix arising in the Pauli Lubanski vector operator.
Entanglement of the eigenvectors and its connection
with degeneracy is discussed.

\strut\hfill

Let $S_1$, $S_2$, $S_3$ be the spin matrices \cite{1} for spin 
$s=\frac12,\, 1,\,\frac32,\, 2,\dots$. 
The matrices are $(2s+1) \times (2s+1)$ hermitian matrices 
($S_1$ and $S_3$ are real symmetric) with trace equal to 0 satisfying 
the commutation relations
$$
[S_1,S_2] = iS_3, \quad [S_2,S_3] = iS_1, \quad [S_3,S_1] = iS_2.
\eqno(1)
$$
The eigenvalues of $S_1$, $S_2$, $S_3$ are $s,s-1,\dots,-s$ for a 
given $s$. We have 
$$
S_1^2 + S_2^2 + S_3^2 = s(s+1)I_{2s+1}
\eqno(2)
$$
where $I_{2s+1}$ is the $(2s+1) \times (2s+1)$ identity matrix.
Furthermore we have
$$
\mbox{tr}(S_j^2)=\frac13 s(s+1)(2s+1)
\eqno(3) 
$$
and
$$
\mbox{tr}(S_jS_k)=0 \quad \mbox{for} \quad j \ne k 
\quad \mbox{and} \, j,k=1,2,3.
\eqno(4)
$$
In general we have that
$\mbox{tr}(S_j^n)=0$ if $n$ is odd and 
$\mbox{tr}(S_j^n)=\frac13 s(s+1)(2s+1)$ if $n$ is even. 
It follows that $\mbox{tr}((S_jS_k) \otimes (S_\ell S_m))=0$
if $j \ne k$ and $\ell \ne m$.
\newline

The Pauli-Lubanski vector operator
(see \cite{2,3,4} and references therein) is given by
$$
W^\mu = 
\sum_{\nu=0}^3\sum_{\rho=0}^3\sum_{\lambda=0}^3 
\epsilon^{\mu\nu\rho\lambda}S_{\rho\lambda}p_{\nu}
$$
where $\mu=0,1,2,3$ and the underlying metric tensor field is
$$
g = -dx_0 \otimes dx_0+dx_1 \otimes dx_1+dx_2 \otimes dx_2
+dx_3 \otimes dx_3.
$$ 
The $W^{\mu}$ ($\mu=0,1,2,3$) are $(2s+1) \times (2s+1)$ matrices
with
$$
\sum_{\mu=0}^3 W_{\mu}W^{\mu} = 
s(s+1)(\sum_{\mu=0}^3 p_{\mu}p^{\mu})I_{2s+1}
$$
and $S=(S_{\rho\lambda})$ denotes the 
$(4\cdot (2s+1)) \times (4\cdot (2s+1))$ matrix
$$
S = (S_{\rho\lambda}) = 
\pmatrix { 0_{2s+1} & S_1 & S_2 & S_3 \cr -S_1 & 0_{2s+1} & -iS_3 & iS_2 \cr
-S_2 & iS_3 & 0_{2s+1} & -iS_1 \cr -S_3 & -iS_2 & iS_1 & 0_{2s+1} }.
$$
Here $0_{2s+1}$ is the $(2s+1) \times (2s+1)$ zero matrix.
The inverse matrix of $S$ is given by
$$
S^{-1} = \frac1{s(s+1)}
\pmatrix { -I_{2s+1} & -S_1 & -S_2 & -S_3 \cr
S_1 & I_{2s+1} & -iS_3 & iS_2 \cr
S_2 & iS_3 & I_{2s+1} & -iS_1 \cr
S_3 & -iS_2 & iS_1 & I_{2s+1} }.
$$
We also write $S=T_1+T_2$ with 
$$
T_1 = 
\pmatrix { 0_{2s+1} & S_1 & S_2 & S_3 \cr -S_1 & 0_{2s+1} & 0_{2s+1} & 0_{2s+1} \cr
-S_2 & 0_{2s+1} & 0_{2s+1} & 0_{2s+1} \cr -S_3 & 0_{2s+1} & 
0_{2s+1} & 0_{2s+1} },
\quad
T_2 = 
\pmatrix { 0_{2s+1} & 0_{2s+1} & 0_{2s+1} & 0_{2s+1} \cr 0_{2s+1} & 0_{2s+1} & -iS_3 & iS_2 \cr
0_{2s+1} & iS_3 & 0_{2s+1} & -iS_1 \cr 0_{2s+1} & -iS_2 & iS_1 & 0_{2s+1} }
$$
where $T_1$ is a skew-hermitian matrix and $T_2$ is a hermitian matrix. 
We study the eigenvalue problem for the matrix $S$ and the 
commutators generated by $T_1$ and $T_2$.
First we note that the matrix $S$ is nonnormal. We obtain
$$
[S,S^*] = SS^*-S^*S = 
2 \pmatrix { 0_{2s+1} & i[S_2,S_3] & i[S_3,S_1] & i[S_1,S_2]
\cr
i[S_2,S_3] & 0_{2s+1} & 0_{2s+1} & 0_{2s+1} \cr
i[S_3,S_1] & 0_{2s+1} & 0_{2s+1} & 0_{2s+1} \cr
i[S_1,S_2] & 0_{2s+1} & 0_{2s+1} & 0_{2s+1} }.
$$
The trace $\mbox{tr}(S^N)$ can also be found in closed form.
We obtain
\begin{eqnarray*}
\mbox{tr}(S) &=& 0 \\ 
\mbox{tr}(S^2) &=& 0 \\
\mbox{tr}(S^{3}) &=& s(s+1)(2s+1) \times 2\\
\mbox{tr}(S^{4}) &=& s(s+1)(2s+1)\times 4 \left(s(s+1)-1\right)\\
\mbox{tr}(S^{5}) &=& s(s+1)(2s+1)\times \big(6-8 s (s+1)\big)\\
\mbox{tr}(S^{6}) &=& s(s+1)(2s+1)\times \big(4 (3 s (s+1)-2)\big)\\
\mbox{tr}(S^{7}) &=& s(s+1)(2s+1)\times 2 \left(s (s+1) \left(s^2+s-8\right)+5\right)\\
\mbox{tr}(S^{8}) &=& s(s+1)(2s+1)\times 4 \left(s^6+3 s^5+s^4-3 s^3+3 s^2+5 s-3\right)
\end{eqnarray*}
Based on the properties of matrices $T_{1}$ and $T_{2}$, it is 
possible to show that the trace of $S^{N}$ is
\begin{eqnarray*}
\mbox{tr}(S^{N}) &=& 
\bigg[\frac12(-1+i\sqrt{4s(s+1)-1}) \bigg]^{N}(2s+1) \\
&& + \bigg[\frac{1}{2}(-1-i\sqrt{4s(s+1)-1})\bigg]^{N}(2s+1)\\
&& + (-1)^{N}(s+1)^{N}(2s-1) + s^{N}(2s+3).
\end{eqnarray*}
The formula above shows the eigenvalues along with their 
multiplicity.
\newline

For spin-$\frac12$ we find the eight eigenvalues
$$
\frac12 \,\,\, (4 \, \mbox{times}), \,\,\,
\frac12(-1+\sqrt2 i) \,\,\, (2 \, \mbox{times}), \,\,\,
\frac12(-1-\sqrt2 i) \,\,\, (2 \, \mbox{times}) 
$$
For spin-1 we find the twelve eigenvalues
$$
1 \,\,\, (5 \, \mbox{times}), \,\,\, 
\frac12(-1+\sqrt7 i) \,\,\, (3 \, \mbox{times}), \,\,\,
\frac12(-1-\sqrt7 i) \,\,\, (3 \, \mbox{times}), \,\,\,
-2 \,\,\, (1 \, \mbox{times})
$$
For spin-$\frac32$ we find sixteen eigenvalues
$$
\frac32 \,\,\, (6 \, \mbox{times}), \,\,\, 
\frac12(-1+\sqrt{14}i) \,\,\, (4 \, \mbox{times}), \,\,\,
\frac12(-1-\sqrt{14}i) \,\,\, (4 \, \mbox{times}), \,\,\,
-\frac52 \,\,\, (2 \, \mbox{times})  
$$
For spin-2 we find the twenty eigenvalues
$$
2 \,\,\, (7 \, \mbox{times}), \,\,\,
\frac12(-1+\sqrt{23}i) \,\,\, (5 \, \mbox{times}), \,\,\,
\frac12(-1-\sqrt{23}i) \,\,\, (5 \, \mbox{times}), \,\,\,
-3 \,\,\, (3 \, \mbox{times}). 
$$
For spin-$\frac52$ we find the twenty four eigenvalues 
$$
\frac52 \,\,\, (8 \, \mbox{times}), \quad
\frac12(-1+\sqrt{34}i) \,\,\, (6 \, \mbox{times}), \quad
\frac12(-1-\sqrt{34}i) \,\,\, (6 \, \mbox{times}), \quad
-\frac72 \,\,\, (4 \, \mbox{times}). 
$$
For the general case with spin-$s$ the eigenvalue
$s$ is $(2s+1)+2=2s+3$ degenerate.
For spin $s \ge 1$ we find the eigenvalue 
$(-s-1)$ which is $(2s+1)-2)=2s-1$ degenerate.
The complex eigenvalues 
$$
\frac12(-1 + i\sqrt{4s(s+1)-1}), \qquad
\frac12(-1 - i\sqrt{4s(s+1)-1})
$$
are $(2s+1)$ times degenrate.
\newline

Consider now the commutators and anti-commutators 
for $T_1$ and $T_2$. We obtain
$$
T_3 := [T_1,T_2] = 
\pmatrix { 0_{2s+1} & -S_1 & -S_2 & -S_3 \cr
-S_1 & 0_{2s+1} & 0_{2s+1} & 0_{2s+1} \cr 
-S_2 & 0_{2s+1} & 0_{2s+1} & 0_{2s+1} \cr
-S_3 & 0_{2s+1} & 0_{2s+1} & 0_{2s+1} }.
$$
Furthermore we find
\begin{eqnarray*}
[T_3,T_1] &=& 
2\pmatrix { 
s(s+1)I_{2s+1} & 0_{2s+1} & 0_{2s+1} & 0_{2s+1} \\
0_{2s+1} & -S^2_1 & -S_1S_2 & -S_1S_3 \\
0_{2s+1} & -S_2S_1 & -S^2_2 & -S_2S_3 \\ 
0_{2s+1} & -S_3S_1 & -S_3S_2 & -S^2_3 } \cr
[T_3,T_2] &=& T_1
\end{eqnarray*}
and
\begin{eqnarray*}
[[T_3,T_1],T_1] &=& -4s(s+1)T_{3} \cr
[[T_3,T_1],T_2] &=& 0_{4(2s+1)} \cr
[[T_3,T_2],T_1] &=& 0_{4(2s+1)} \cr
[T_3,T_2],T_2] &=& T_3.  
\end{eqnarray*}
The anticommutator of $T_1$ and $T_2$ is $[T_1,T_2]_+=-T_1$.
We note that
$T_{1}$ and $T_{2}$ are both separately normal with
$T^*_1=-T_1$ $T^*_2=T_2$ and $[S,S^*]=2T_3$.
\newline

For spin-$\frac12$ and eigenvalue $\frac12$ (four times
degenerate) the set of normalized linearly independent 
eigenvectors is given by
$$
{\bf v}_1 = \frac1{\sqrt2} 
\pmatrix { 0 \cr 0 \cr 1 \cr 0 \cr 0 \cr 0 \cr 0 \cr 1 } =
\frac1{\sqrt2}\left(
\pmatrix { 1 \cr 0 } \otimes \pmatrix { 0 \cr 1 } \otimes 
\pmatrix { 1 \cr 0 } + 
\pmatrix { 0 \cr 1 } \otimes \pmatrix { 0 \cr 1 } 
\otimes \pmatrix { 0 \cr 1 }\right) 
$$
$$
{\bf v}_2 = \frac1{\sqrt2} 
\pmatrix { 0 \cr 0 \cr 0 \cr 0 \cr 1 \cr 0 \cr 0 \cr i } =
\pmatrix { 0 \cr 1 } \otimes 
\frac1{\sqrt2}\pmatrix { 1 \cr 0 \cr 0 \cr i } 
$$
$$
{\bf v}_3 = \frac1{\sqrt2} 
\pmatrix { 0 \cr 0 \cr 0 \cr 1 \cr 0 \cr 0 \cr -1 \cr 0 } =
\frac1{\sqrt2}\left(\pmatrix { 1 \cr 0 } \otimes \pmatrix { 0 \cr 1 } 
\otimes \pmatrix { 0 \cr 1 } - 
\pmatrix { 0 \cr 1 } \otimes \pmatrix { 0 \cr 1 } 
\otimes  \pmatrix { 1 \cr 0 }\right) 
$$
$$
{\bf v}_4 =  
\frac1{\sqrt2} \pmatrix { 0 \cr 0 \cr 0 \cr 0 \cr 0 \cr 1 \cr i \cr 0 } =
\pmatrix { 0 \cr 1 } \otimes 
\frac1{\sqrt2} \pmatrix { 0 \cr 1 \cr i \cr 0 }. 
$$
Applying the $n$-tangle introduced by Wong and Christensen 
(\cite{5,6}) we find that the four eigenvectors are non-entangled.
For the second and fourth this also seen from the fact
that the eigenvectors can be written as Kronecker products
of normalized vectors in ${\mathbb C}^2$ and ${\mathbb C}^4$. 
Now forming combinations of these vectors (which are also 
then eigenvectors) we find the following, where
we normalize the linear combinations.
${\bf v}_1+{\bf v}_3$ and ${\bf v}_1-{\bf v}_3$
are non-entangled as well as ${\bf v}_2+{\bf v}_4$
and ${\bf v}_2-{\bf v}_4$ are non-entangled.
On the other hand ${\bf v}_1+{\bf v}_4$, 
${\bf v}_1-{\bf v}_4$, ${\bf v}_2+{\bf v}_3$,
${\bf v}_2-{\bf v}_3$ are entangled.

\end{document}